\documentclass[11pt]{article}

\usepackage[final]{acl}

\usepackage{times}
\usepackage{latexsym}

\usepackage[T1]{fontenc}

\usepackage[utf8]{inputenc}

\usepackage{microtype}

\usepackage{inconsolata}

\usepackage{graphicx}

\usepackage{enumitem}
\usepackage{hyperref}
\usepackage{url}
\usepackage{booktabs} %
\usepackage{amsmath}
\usepackage{amssymb}
\usepackage{amsthm}
\usepackage{multirow}
\usepackage{multicol}
\usepackage{xspace}
\usepackage[table]{xcolor} %
\usepackage{bm}
\usepackage{amsfonts}
\usepackage{makecell}
\usepackage{subfigure}
\usepackage[noabbrev,capitalise]{cleveref}
\usepackage{tcolorbox}
\tcbuselibrary{breakable,skins}

\newcounter{prompt}

\newtcolorbox{promptbox}[2][]{
  breakable,
  enhanced,
  colback=gray!5,
  colframe=gray!30,
  boxrule=0.4pt,
  arc=2pt,
  left=6pt, right=6pt, top=4pt, bottom=4pt,
  fontupper=\ttfamily\small,
  title={#2},
  coltitle=black,
  fonttitle=\normalfont\small,
  attach boxed title to top left={yshift=-2pt, xshift=2pt},
  boxed title style={
    colback=gray!15,
    boxrule=0pt,
    arc=2pt,
    left=4pt, right=4pt,
    top=1pt, bottom=1pt
  },
  minipage boxed title=0.91\linewidth,  %
  before upper={\refstepcounter{prompt}\ifx&#1&\else\label{#1}\fi},
  before=\vspace{3pt}, after=\vspace{3pt},
  nameref={#2}
}

\crefname{prompt}{Prompt}{Prompts}
\Crefname{prompt}{Prompt}{Prompts}

\newcommand{\eg}{\emph{e.g.,}\xspace}

\newcommand{\wrt}{w.r.t.\xspace}
\newcommand{\ignore}[1]{}

\newcommand{\NA}{\textit{n/a}}

\title{Bridging Language and Items for Retrieval and Recommendation:\\ Benchmarking LLMs as Semantic Encoders}

\author{
        \textbf{Yupeng Hou\textsuperscript{$1\dagger$}},
        \textbf{Jiacheng Li\textsuperscript{$1\dagger$}}, 
        \textbf{Xiangjun Fu\textsuperscript{$1\dagger$}}, 
        \\
        \textbf{Zhankui He}$^1$,
        \textbf{An Yan}$^1$,
        \textbf{Xiusi Chen}$^2$,
        \textbf{Julian McAuley\textsuperscript{$1$}}
        \vspace{2mm}
       \\ 
  \textsuperscript{1}UC San Diego,
  \textsuperscript{2}UC Los Angeles
  \\
  {\tt \{yphou,j9li,xif001,zhh004,ayan,jmcauley\}@ucsd.edu\ \ xchen@cs.ucla.edu}
  \vspace{2mm}
  \\
  Amazon Reviews 2023 dataset: \url{https://amazon-reviews-2023.github.io}
  \\
  Benchmarking toolkit: \url{https://github.com/hyp1231/BLaIR-Bench}
}

\begin{document}
\maketitle
\begin{abstract}
Feature engineering has long been central to recommender systems, yet effectively leveraging textual item features remains challenging. Recent advances in large language models (LLMs) have enabled their use as semantic encoders for recommendation, but their roles and behaviors in this setting are still not well understood. Prior studies often rely on general-purpose embedding benchmarks (\eg MTEB) when selecting LLMs, overlooking the unique characteristics of recommendation tasks. To address this gap, we introduce \textsc{BLaIR}, a comprehensive benchmark for evaluating LLMs as semantic encoders in recommendation scenarios. We contribute (1) a new large-scale Amazon Reviews 2023 dataset with over 570 million reviews and 48 million items, (2) a unified benchmark covering sequential recommendation, collaborative filtering, and product search, and (3) a new complex-query product search task featuring both semi-synthetic and real-world evaluation datasets. Experiments with 11 leading LLMs show that their rankings on \textsc{BLaIR} show little correlation with MTEB, highlighting the unique challenges of semantic encoding in recommendation.
\end{abstract}

\section{Introduction}

Feature engineering has long been central to modern recommender systems~\cite{rendle2010factorization,chen2012svdfeature,cheng2016wide}. However, effectively leveraging textual information (\eg item descriptions) has remained a challenge. While text features are rich in semantics, they are often noisy, unstructured, and difficult to integrate directly into recommendation models. Early approaches relied on keyword tagging~\cite{mooney2000content,wang2011collaborative,tang2012cross}, which were largely heuristic and discarded substantial semantic information. Moreover, textual features were typically modeled using shallow statistical techniques, lacking the broader common-sense knowledge inherent in natural language.

\begin{figure*}[!t]
  \centering
  \includegraphics[width=0.85\linewidth]{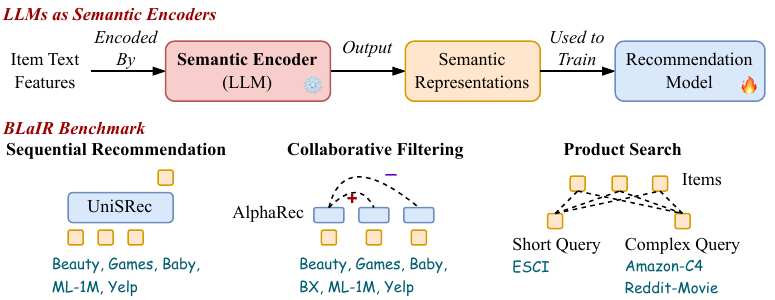}
  \caption{Overview of \textsc{BLaIR} benchmark. We evaluate LLMs as semantic encoders to convert item text features into semantic representations, which then train downstream recommendation models. The benchmark covers three scenarios: (i) sequential recommendation (\eg UniSRec~\cite{hou2022unisrec}); (ii) collaborative filtering (\eg AlphaRec~\cite{sheng2024language}); and (iii) product search with short queries and complex queries.}
  \label{fig:overview}
\end{figure*}

With the rapid advancement of large language models (LLMs)~\cite{hoffmann2022training,rae2021scaling,chowdhery2023palm,touvron2023llama,zhao2023llm_survey}, researchers have begun to employ them as semantic encoders for recommendation. Pretrained on large-scale corpora, LLMs capture extensive world knowledge and demonstrate strong capabilities in understanding and representing natural language. Consequently, an increasing number of studies has explored using LLM-encoded text representations as inputs to train recommendation models, a direction referred to as modality-based recommendation~\cite{hou2022unisrec,yuan2023go}. This paradigm has shown promising results across various tasks, including sequential recommendation~\cite{hou2022unisrec,yuan2023go}, collaborative filtering~\cite{ren2024representation,sheng2024language}, and product search~\cite{li2021embedding,reddy2022esci}.

Despite this progress, the role of LLMs as semantic encoders in recommendation remains not well understood. Existing studies typically adopt popular or recently released LLMs, or rely on general-purpose sentence embedding benchmarks such as MTEB~\cite{muennighoff2023mteb,enevoldsen2025mmteb} for model selection. However, the requirements of recommendation differ substantially from those of generic text embedding tasks.
(1) In recommendation, encoded representations are usually used as input features to train downstream models such as Transformer decoders for sequential recommendation, whereas in text embedding tasks, representations are more often consumed directly for similarity search or simple classifiers.
(2) In recommendation, text inputs are often short and noisy (\eg item titles), requiring semantic encoders to provide strong disambiguation based on world knowledge, while text embedding benchmarks typically evaluate well-formed sentences or paragraphs.

To address this gap, we present \textsc{BLaIR}, which stands for \textbf{B}ridging \textbf{La}nguage and \textbf{I}tems for \textbf{R}etrieval and \textbf{R}ecommendation, a comprehensive benchmark designed to evaluate LLMs as semantic encoders in recommendation scenarios (\Cref{fig:overview}). Our contributions are as follows:\\
$\bullet $ \textbf{A new large-scale e-commerce dataset: Amazon Reviews 2023.} Previous Amazon Reviews datasets~\cite{mcauley2015image,he2016ups,ni2019justifying} are widely used for benchmarking but were released several years ago (last released in 2018) and contain noisy metadata. 
We collect a new dataset with cleaned metadata, containing over 570 million reviews and 48 million items, with a cutoff date of September 2023.\\
$\bullet $ \textbf{A comprehensive benchmark: \textsc{BLaIR}.} We evaluate three representative recommendation scenarios: sequential recommendation, collaborative filtering, and product search. For each, we implement strong baselines and conduct extensive experiments with 11 top-performing LLMs on MTEB.\\
$\bullet $ \textbf{A new subtask: complex-query product search.} Existing product search datasets typically involve short queries. To align with emerging applications that require understanding long, complex, and ambiguous queries, we introduce a new subtask, complex-query product search. We construct two datasets: one semi-synthetic dataset derived from real user reviews and another curated from real-world forum posts.

Our results yield several notable insights:
(1) LLM rankings on \textsc{BLaIR} are not positively correlated with their rankings on MTEB, highlighting that recommendation poses distinct challenges compared to general embedding benchmarks.
(2) Even when downstream models have the same number of parameters, performance still scales with the size of the semantic encoder, though the effect weakens as task complexity grows.
(3) We observe a strong correlation between results on semi-synthetic and real-world datasets for complex query product search tasks, implying that semi-synthetic data can be a cost-effective proxy for evaluation.

\section{Related Work}

\paragraph{Semantic Encoders in Recommendation.} Researchers have long sought to better leverage semantically rich textual features such as titles and descriptions. Early works in content-based recommendation~\cite{mooney2000content,gopalan2014content} extracted keywords and learned their collaborative relations between users and items~\cite{rendle2010factorization,wang2011collaborative,chen2012svdfeature,wang2015collaborative}. With advances in natural language processing (NLP), later studies began directly encoding text features using pretrained models to better capture language semantics~\cite{chen2017joint,hou2022unisrec,yuan2023go}. However, such models often suffer from domain gaps between their pretraining data and recommendation scenarios~\cite{hou2023vqrec}. Recently, the emergence of LLMs with superior generalization capabilities has inspired researchers to explore using LLMs as semantic encoders in recommendation~\cite{wei2024llmrec,ren2024representation,ren2024easyrec,sheng2024language}. 
Nevertheless, questions remain about how the choice of LLM impacts the downstream performance. 

\paragraph{Benchmarking Text Embeddings.} To comprehensively evaluate text embedding models across domains and tasks, researchers have developed benchmarks that aggregate multiple public datasets. An early example is BEIR~\cite{thakur2021beir}, which evaluates text embedding models on information retrieval tasks. MTEB~\cite{muennighoff2023mteb} and MMTEB~\cite{enevoldsen2025mmteb} further expand the evaluation scope beyond retrieval to broader embedding applications such as classification and clustering. Recently, BRIGHT~\cite{su2025bright} focuses on retrieval tasks that require reasoning to identify relevant documents. In this work, we aim to benchmark text embedding models as semantic encoders in recommendation scenarios. This is not merely a domain-specific adaptation but rather reflects that, as semantic encoders, LLMs 
require different capabilities.

\paragraph{Benchmarking LLMs for Recommendation.} Existing efforts to benchmark LLMs for recommendation can be grouped into three categories: (1) benchmarking their real-world web interaction capabilities as shopping agents~\cite{yao2022webshop,jiang2025shoppingbench}; (2) benchmarking their shopping-related knowledge and understanding of user intent~\cite{jin2024shoppingmmlu}; and (3) benchmarking LLMs themselves as recommendation models~\cite{liu2023llmrec,liu2024RSBench,jiang2025beyond,liu2025benchmarking}. In contrast, to the best of our knowledge, we are the first to benchmark LLMs for encoding textual features in recommendation scenarios.

\begin{table}[t]
\small
\centering
\caption{Statistics of different versions of Amazon Reviews datasets. \#Tokens denotes the total number of text tokens obtained using the \texttt{cl100k\_base} tokenizer.}
\label{tab:data_stat}
\resizebox{0.95\columnwidth}{!}{
\setlength{\tabcolsep}{1mm}{
\begin{tabular}{lrrrr}
\toprule
\textbf{Version} & \textbf{Amazon'13} & \textbf{Amazon'14} & \textbf{Amazon'18} & \textbf{Amazon'23} \\ 
\midrule
\#Categories     & 28                 & 24                 & 29                 & \textbf{33}        \\ 
\midrule
\rowcolor{gray!20}\multicolumn{5}{c}{\textbf{\textit{Customer Reviews}}} \\ 
\midrule
\#Reviews        & 34,686,771         & 82,456,877         & 233,055,327        & \textbf{571,544,897} \\
\#Users          & 6,643,669          & 21,128,805         & 43,531,850         & \textbf{54,514,264}  \\
\#Items          & 2,441,053          & 9,857,241          & 15,167,257         & \textbf{48,185,153}  \\
\#Tokens         & 5.9B               & 9.1B               & 15.7B              & \textbf{30.1B}       \\
Min Time         & Jun 1995           & Jun 1996           & Jun 1996           & Jun 1996             \\
Max Time         & Mar 2013           & July 2014          & Oct 2018           & \textbf{Sep 2023}    \\ 
\midrule
\rowcolor{gray!20}\multicolumn{5}{c}{\textbf{\textit{Item Metadata}}} \\ 
\midrule
\#Tokens         & -                  & 4.1B               & 7.9B               & \textbf{30.7B}       \\
\# Meta          & -                  & 9,430,088          & 14,741,571         & \textbf{35,393,189}  \\ 
\bottomrule
\end{tabular}
}
}
\end{table}

\begin{table}
        \small
    \centering
    \caption{Statistics of representative large-scale recommendation datasets.}\label{tab:dataset}
    \resizebox{\columnwidth}{!}{
    \setlength{\tabcolsep}{0.8mm}{
    \begin{tabular}{lrrrr}
    \toprule
    \textbf{Dataset} & \textbf{\#Items} & \textbf{\#Users} & \textbf{\#Interactions} \\
    \midrule
    Netflix~\cite{bennett2007netflix} & 17,770 & 480,189 & 100,480,507  \\
    Amazon'18~\cite{ni2019justifying} & 15,167,257 & 43,531,850 & 233,055,327 \\
    Google Local~\cite{yan2023personalized} & 4,963,111 & 113,643,107 & 666,324,103  \\
    Tenrec~\cite{yuan2022tenrec} & 3,753,436 & 5,022,750 & 142,321,193 \\
    MicroLens~\cite{ni2023content} & 1,142,528 &  34,492,051 &  1,006,528,709  \\ \midrule
    Amazon Reviews 2023 & 48,185,153 & 54,514,264 & 571,544,897  \\
    \bottomrule
    \end{tabular}
    }}
\end{table}

\section{A Large-Scale E-Commerce Dataset: Amazon Reviews 2023}\label{sec:amazon}

The Amazon Reviews datasets are among the most representative publicly available resources for evaluating recommendation models~\cite{mcauley2015image,he2016ups,ni2019justifying}, offering rich item features. Given the wide use of the Amazon Reviews datasets, we aim to include them in our benchmark. However, existing versions of the Amazon Reviews datasets are outdated. The most recent one was last updated in 2018.

\begin{table*}[t]
\small
\centering
\caption{Data statistics of the \textsc{BLaIR} benchmark. Each row is a dataset. Cells that do not apply are marked \NA. ``Real'' indicates real-world data, while ``Sync'' indicates semi-synthetic data. More details about data split strategies (by timestamp, by ratio, or leave-last-out) can be found in~\Cref{app:data-split}.}
\label{tab:all-data-stats}
\scriptsize
\resizebox{\linewidth}{!}{
\setlength{\tabcolsep}{1mm}{
\begin{tabular}{
    l                %
    l                %
    l                %
    r                %
    r r r            %
    c                %
    r                %
    r                %
    r                %
}
\toprule
\multirow[c]{2.5}{*}{\textbf{Dataset}} &
\multirow[c]{2.5}{*}{\makecell[c]{\textbf{Data Source}\\\textbf{/ Subtask}}} &
\multirow[c]{2.5}{*}{\textbf{Type}} &
\multirow[c]{2.5}{*}{\textbf{\#Items}} &
\multicolumn{3}{c}{\textbf{\#Interactions}} &
\multirow[c]{2.5}{*}{\textbf{Split}} &
\multirow[c]{2.5}{*}{\makecell[c]{\textbf{Avg. \#Items}\\\textbf{Per User}}} &
\multirow[c]{2.5}{*}{\makecell[c]{\textbf{Avg. Item}\\\textbf{Metadata \#Chars}}} &
\multirow[c]{2.5}{*}{\makecell[c]{\textbf{Avg. Query}\\\textbf{\#Chars}}}
\\
\cmidrule(lr){5-7}
 & & & & \multicolumn{1}{c}{\textbf{Train}} & \multicolumn{1}{c}{\textbf{Val}} & \multicolumn{1}{c}{\textbf{Test}} & \multicolumn{4}{c}{} \\
\midrule
\rowcolor{gray!20}\multicolumn{11}{c}{\textbf{\textit{Sequential Recommendation}}}\\
\midrule
\textbf{Beauty} & \multirow[c]{3}{*}{Amazon'23} & \multirow[c]{5.5}{*}{Real} & 43,978 & 82,800 & 14,268 & 7,698 & \multirow[c]{3}{*}{\makecell[c]{By Timestamp\\(08/11/2021,\\07/16/2022)}} & 2.45 & 118.83 & \multirow[c]{5.5}{*}{\NA} \\
\textbf{Games} & & & 115,813 & 2,127,563 & 189,268 & 215,809 & & 3.41 & 85.40 \\
\textbf{Baby} & & & 179,128 & 2,930,822 & 304,617 & 347,884 & & 3.53 & 107.26 \\
\addlinespace
\textbf{ML-1M} & ML-1M & & 3,043 & 983,412 & 6,040 & 6,040 & \multirow[c]{2}{*}{\makecell[c]{Leave-\\Last-Out}} & 164.82 & 23.78 \\
\textbf{Yelp} & Yelp & & 20,033 & 255,492 & 30,431 & 30,431 & & 10.01 & 604.60 \\
\midrule
\rowcolor{gray!20}\multicolumn{11}{c}{\textbf{\textit{Collaborative Filtering}}}\\
\midrule
\textbf{Beauty} & \multirow[c]{3}{*}{Amazon'23} & \multirow[c]{6.5}{*}{Real} & 43,978 & 41,906 & 31,429 & 31,431 & \multirow[c]{6.5}{*}{\makecell[c]{By Ratio\\(4:3:3)}} & 2.45 & 118.83 & \multirow[c]{6.5}{*}{\NA}\\
\textbf{Games} & & & 115,813 & 1,013,056 & 759,792 & 759,792 & & 3.41 & 85.40 \\
\textbf{Baby} & & & 179,128 & 1,433,329 & 1,074,996 & 1,074,998 & & 3.53 & 107.26 \\
\addlinespace
\textbf{BX} & BX & & 5,335 & 101,183 & 75,804 & 75,862 & & 40.31 & 27.57 \\
\textbf{ML-1M} & ML-1M & & 3,043 & 398,184 & 298,631 & 298,677 & & 164.82 & 23.78 \\
\textbf{Yelp} & Yelp & & 20,033 & 126,542 & 94,906 & 94,906 & & 10.01 & 604.60 \\
\midrule
\rowcolor{gray!20}\multicolumn{11}{c}{\textbf{\textit{Product Search}}}\\
\midrule
\textbf{ESCI} & Short & Real & 1,367,729 & \multirow[c]{3}{*}{\NA}& \multirow[c]{3}{*}{\NA} & 27,643 & \multirow[c]{3}{*}{\NA} & \multirow[c]{3}{*}{\NA} & 96.57 & 22.46 \\
\textbf{Amazon-C4} & Complex & Sync & 1,058,417 & & & 21,223 & & & 99.81 & 229.89 \\
\textbf{R-Movies} & Complex & Real & 51,203 & & & 1,627 & & & 19.63 & 435.85 \\
\bottomrule
\end{tabular}
}
}
\end{table*}

To enhance the quality of our proposed benchmark, we collect a new version of the dataset, named \emph{Amazon Reviews 2023}\footnote{\url{https://amazon-reviews-2023.github.io/}}. \Cref{tab:data_stat}~presents key statistics comparing Amazon Reviews 2023 with earlier versions, while \Cref{tab:dataset}~compares Amazon Reviews 2023 with other representative large-scale datasets. The new dataset offers several notable improvements:
\begin{itemize}[nosep,leftmargin=*]
\item \emph{Larger scale}: Amazon Reviews 2023 is substantially larger than previous versions across all dimensions. Specifically, it contains $3.18\times$ more items and $2.58\times$ more text tokens in reviews and item metadata than the 2018 version.
\item \emph{More recent interactions}: The dataset contains more recent reviews from Amazon, extending the previous version with new data ranging from Oct 2018 to Sep 2023.
\item \emph{Richer and cleaner metadata}: We re-parsed the original HTML product pages into structured JSON format, resulting in metadata with more descriptive fields (\eg item descriptions and features) and multi-modal information (\eg product videos and images at multiple resolutions).
\item \emph{Finer-grained timestamps}: Previous datasets provided timestamps only at the day level, which can introduce inaccuracies for time-sensitive tasks such as sequential recommendation. In contrast, Amazon Reviews 2023 offers timestamps with millisecond precision.
\end{itemize}

\section{\textsc{BLaIR} Benchmark}

In this section, we present the structure of the proposed benchmark, \textsc{BLaIR}. The benchmark comprises three recommendation-related scenarios: sequential recommendation, collaborative filtering, and product search. We first introduce the definitions and used datasets of each task in~\Cref{sec:task_def}. Then, we describe in detail a newly introduced subtask under the product search scenario, namely complex-query product search, which contains long and ambiguous user queries (\Cref{sec:new_task}).

\subsection{Task Definitions and Datasets}\label{sec:task_def}

Given an item $v$ with textual features $x_v$ (such as title), the first step in using LLMs as semantic encoders is to transform these text features into dense representations: $\bm{e}_v = \text{LLM}(x_v)$. Here, $\text{LLM}(\cdot)$ denotes a pretrained and frozen LLM, and $\bm{e}_v \in \mathbb{R}^d$ is the resulting semantic representation for item $v$. The encoded representations $\{\bm{e}_v\}$ can then be used as inputs to train downstream recommendation models or directly for retrieval tasks. We consider three representative scenarios, each with its own task definition and dataset. The statistics of all datasets used in \textsc{BLaIR} are summarized in~\Cref{tab:all-data-stats}.

\paragraph{Sequential Recommendation.} In the sequential recommendation task, a user's interaction history is represented as a time-ordered sequence of items $S_{t-1} = [v_1, v_2, \ldots, v_{t-1}]$. The goal is to predict the next item $v_t$ that the user will interact with. Following common practice~\cite{hou2022unisrec,yuan2023go,sheng2024language}, we first add adapter layers to align the semantic embedding space with the recommendation space. Specifically, each item embedding is projected as $\bm{e}'_v = \text{Proj}_{\text{seq}}(\bm{e}_v) \in \mathbb{R}^{d'}$, where $\text{Proj}_{\text{seq}}(\cdot)$ denotes a learnable projection layer. This design ensures that the downstream recommendation model maintains a consistent number of parameters, regardless of the dimensionality of the underlying LLM.
We then adopt the UniSRec~\cite{hou2022unisrec} architecture to implement the sequential recommendation model. Formally, the next-item probability is defined as:
\begin{equation*}
P(v_t | S_{t-1}) \propto \text{Trm}(\bm{e}'_{v_1}, \bm{e}'_{v_2}, \ldots, \bm{e}'_{v_{t-1}}) \cdot \bm{e}'_{v_t},
\end{equation*}
where $\text{Trm}(\cdot)$ is a Transformer decoder~\cite{vaswani2017attention}. The model is trained by maximizing the likelihood of the next items using a cross-entropy loss.

For sequential recommendation tasks, we select three categories from the Amazon Reviews 2023 dataset: All Beauty (\emph{Beauty}), Video Games (\emph{Games}), and Baby Products (\emph{Baby}). These categories differ in both scale (0.7M, 4.5M, and 6.0M interactions, respectively) and domain (fashion, entertainment, and childcare). In addition, we include two widely used public datasets, Movielens-1M (\emph{ML-1M})~\cite{harper2015movielens} and \emph{Yelp}~\cite{yelp_dataset}.

\paragraph{Collaborative Filtering.} Given historical user-item interactions, the goal of collaborative filtering is to predict which item a user $u$ is most likely to interact with next. Similar to the sequential recommendation setup, we add an adapter layer to obtain projected item representations $\bm{e}'_v = \text{Proj}_{\text{cf}}(\bm{e}_v) \in \mathbb{R}^{d'}$. Following AlphaRec~\cite{sheng2024language}, the user representation $\bm{e}'_u$ is computed as the average of the projected representations of items that the user has interacted with. The prediction probability is then defined as:
\begin{equation*}
P(v | u) \propto \cos(\bm{W}\bm{e}'_u, \bm{W}\bm{e}'_v) / \tau,
\end{equation*}
where $\bm{W}\in\mathbb{R}^{d'' \times d'}$ is a linear layer, $\cos(\cdot,\cdot)$ denotes the cosine similarity function, and $\tau$ is a temperature hyperparameter. The model is trained using the InfoNCE loss with in-batch negative sampling~\cite{chen2020simclr}.

For collaborative filtering tasks, we use the same five datasets as in the sequential recommendation scenario: \emph{Beauty}, \emph{Games}, and \emph{Baby} from Amazon Reviews 2023, along with \emph{ML-1M} and \emph{Yelp}. Additionally, we include the widely used Book-Crossing dataset (\emph{BX})~\cite{ziegler2005bx}. We exclude Book-Crossing from the sequential recommendation scenario because it does not provide timestamps to order user interactions.

\paragraph{Product Search.} In the product search task, given a user query $q$, the goal is to retrieve the most relevant items from a large corpus. The query is encoded using the same LLM as for items: $\bm{e}_q = \text{LLM}(q) \in \mathbb{R}^{d}$. For this task, we do not apply any adapter layers and directly use the raw LLM-encoded representations for retrieval. The relevance score between a query $q$ and an item $v$ is computed as $\bm{e}_q \cdot \bm{e}_v$ in a zero-shot manner.

We consider two subtasks under the product search scenario: (1) \emph{short-query product search}, the conventional setting where user queries are short and specific, typically containing concise keywords; and (2) \emph{complex-query product search}, where user queries are longer, more descriptive, and often ambiguous. For short-query product search, we use the \emph{ESCI} dataset~\cite{reddy2022esci}. The latter subtask is newly introduced in this work, with detailed task definitions and evaluation datasets presented in~\Cref{sec:new_task}.

\subsection{New Subtask: Complex-Query Product Search}\label{sec:new_task}

As the natural language understanding capabilities of large language models continue to grow, users are no longer limited to issuing short, keyword-based queries when searching for items. Instead, they can describe their needs in more natural and expressive ways, which can be long, ambiguous, or require reasoning to identify relevant items (\eg Buy it in ChatGPT\footnote{\url{https://openai.com/index/buy-it-in-chatgpt/}} and Amazon Rufus\footnote{\url{https://www.amazon.com/rufus/}}). To reflect this emerging trend, we introduce this subtask into our benchmark.

However, there are few public datasets that support evaluation in this setting. Collecting real paired complex user queries and their corresponding items is infeasible, as user chat histories are and should remain private. Therefore, we construct two datasets for this task: a semi-synthetic dataset derived from real user reviews in the Amazon Reviews 2023 (collected in~\Cref{sec:amazon}), and a real-world dataset curated from Reddit forum posts.

\begin{table}
        \small
    \centering
    \caption{Comparison of user queries for the same item in the ESCI and Amazon-C4 datasets.}
    \label{tab:case}
    \setlength{\tabcolsep}{1mm}{
    \begin{tabular}{ll}
    \toprule
    \textbf{Item ASIN/ID} & B07RJZNY5C\\
    \addlinespace
    \textbf{Query}~(ESCI) & salt gun \\
    \addlinespace
    \makecell[l]{\textbf{Query}\\ (Amazon-C4)\\ \\ \\ } &  \makecell[l]{I want a gun that I can use while garden-\\ ing to get rid of stink bugs, ants, flies,\\ and spiders in my house. It needs to be\\ amazing and help me feel less scared.} \\
    \addlinespace
    \textbf{Metadata} & BUG-A-SALT 3.0 Black Fly Edition.\\
    \bottomrule
    \end{tabular}
    }
\end{table}

\begin{table}[!t]
    \small
    \centering
    \caption{An example from the Reddit-Movie dataset.}
    \label{tab:reddit_case}
    \setlength{\tabcolsep}{1mm}{
    \begin{tabular}{ll}
    \toprule
    \textbf{IMDb ID} & tt2582802\\
    \addlinespace
    \makecell[l]{\textbf{Query}\\ (Reddit-Movie)\\ \\ \\ \\ \\ \\ \\ } & 
    \makecell[l]{Movies with a character that is very deter-\\mined and goes through a vigorous train-\\ing to be very good/the best at something?\\It doesn't necessarily have to be a train-\\ing, it could be the character learning by\\trial and error, experience and doing the\\thing himself. Pretty much any genre is\\okay, action, sci fi, thriller, horror, ...} \\
    \addlinespace
    \textbf{Movie} & Whiplash (2014) \\
    \bottomrule
    \end{tabular}
    }
\end{table}

\begin{table*}[t]
\small
\centering
\caption{Performance comparison of LLMs as semantic encoders across different recommendation scenarios. ``Seq. Rec.'', ``Col. Fil.'', ``Short'', and ``Complex'' stand for sequential recommendation (5 datasets), collaborative filtering (6 datasets), short-query product search (1 dataset), and complex-query product search (2 datasets), respectively.
The best results in each column are highlighted in \textbf{bold}.}
\label{tab:main-results}
\scriptsize
\resizebox{\linewidth}{!}{
\setlength{\tabcolsep}{1mm}{
\begin{tabular}{llrrrr@{\hspace{4mm}}rrr}
\toprule
\multirow{2.5}{*}{\textbf{Model}} & \multirow{2.5}{*}{\textbf{Rank (Borda)}$^*$} & \multirow{2.5}{*}{\makecell[c]{\textbf{Avg.}\\(Overall)}} & \multirow{2.5}{*}{\makecell[c]{\textbf{Avg.}\\(Task)}} & \multirow{2.5}{*}{\textbf{Seq. Rec.}} & \multirow{2.5}{*}{\textbf{Col. Fil.}} & \multicolumn{2}{c}{\textbf{Product Search}} \\
\cmidrule(lr){7-8}
 &  &  &  &  &  & \textbf{Short} & \textbf{Complex} \\
\midrule
\rowcolor{gray!20}\multicolumn{8}{c}{\textbf{\textit{Open-Source Models (< 1B parameters)}}} \\
\midrule
\texttt{FacebookAI/roberta-large} & 11 (15.0) & 0.0263 & 0.0190 & 0.0393 & 0.0269 & 0.0096 & 0.0001 \\
\texttt{Qwen/Qwen3-Embedding-0.6B} & 10 (35.5) & 0.0507 & 0.0829 & 0.0415 & 0.0274 & 0.1876 & 0.0750 \\
\texttt{princeton-nlp/sup-simcse-roberta-large} & 9 (37.0) & 0.0397 & 0.0536 & 0.0426 & 0.0265 & 0.1063 & 0.0389 \\
\texttt{sentence-transformers/sentence-t5-large} & 8 (42.5) & 0.0513 & 0.0801 & 0.0418 & 0.0304 & 0.1691 & 0.0790 \\
\midrule
\rowcolor{gray!20}\multicolumn{8}{c}{\textbf{\textit{Open-Source Models ($\ge$ 1B parameters)}}} \\
\midrule
\texttt{Qwen/Qwen3-Embedding-4B} & 7 (69.5) & 0.0620 & 0.1036 & 0.0416 & 0.0350 & 0.2258 & 0.1120 \\
\texttt{Qwen/Qwen3-Embedding-8B} & 6 (54.0) & 0.0637 & 0.1069 & 0.0415 & 0.0362 & 0.2328 & 0.1172 \\
\texttt{Salesforce/SFR-Embedding-Mistral} & 4 (98) & 0.0679 & 0.1160 & 0.0433 & 0.0372 & \textbf{0.2560} & 0.1273 \\
\texttt{intfloat/e5-mistral-7b-instruct} & 3 (101) & 0.0666 & 0.1120 & 0.0434 & 0.0377 & 0.2437 & 0.1232 \\
\texttt{GritLM/GritLM-7B} & 2 (105) & \textbf{0.0685} & \textbf{0.1161} & 0.0434 & \textbf{0.0385} & 0.2537 & \textbf{0.1290} \\
\midrule
\rowcolor{gray!20}\multicolumn{8}{c}{\textbf{\textit{Proprietary Models}}} \\
\midrule
\texttt{gemini-embedding-001} (Google) & 5 (96.5) & 0.0629 & 0.1040 & 0.0434 & 0.0355 & 0.2233 & 0.1140 \\
\texttt{text-embedding-3-large} (OpenAI) & \textbf{1 (116)} & 0.0665 & 0.1112 & \textbf{0.0440} & 0.0366 & 0.2366 & 0.1278 \\
\bottomrule
\end{tabular}
}
}
\footnotesize
$^\ast$ Rank (\(\downarrow\)) is determined by the Borda Count score. 
For the Borda score and all other metrics, higher values are better (\(\uparrow\)).
\end{table*}

\paragraph{Semi-Synthetic Complex Queries: Amazon-C4.} The first dataset we introduce is \emph{Amazon-C4}\footnote{\url{https://huggingface.co/datasets/McAuley-Lab/Amazon-C4}}, short for \textbf{C}omplex \textbf{C}ontexts \textbf{C}reated by \textbf{C}hatGPT. The key idea is that user reviews often contain detailed intentions and rich contextual information, which can be repurposed into complex user queries. However, directly using reviews as queries is problematic. Reviews are usually written in a reflective tone, not in the style of a user query. In addition, reviews may include information about the target item that causes data leakage. To address this, we use large language models to rephrase user reviews into natural, query-like forms, creating the semi-synthetic dataset.

\begin{promptbox}[prompt:amazon-c4]{Prompt for complex query synthesizing in Amazon-C4.}
Given a review from Amazon, can you rephrase it with a first-person tone as if you are the customer looking for a product? Give me the rephrased output without saying anything else. Note that the name of the product must not show in the output since you are looking for it. You should ignore irrelevant information that doesn't help with the rewriting.
\end{promptbox}

Specifically, we uniformly sample more than 20,000 5-star reviews from the Amazon Reviews 2023 dataset with at least 100 characters per review to ensure sufficient detail. We use ChatGPT to rephrase each review into a first-person query expressing the intent to find a suitable item (see~\Cref{prompt:amazon-c4} for the prompt). We then filter out cases where ChatGPT fails to perform the conversion. A comparison between a user query in Amazon-C4 and ESCI for the same target item is shown in~\Cref{tab:case}.

\paragraph{Real-World Complex Queries: Reddit-Movies.} The second dataset, Reddit-Movies (\emph{R-Movies}), is curated from real forum discussions. As Amazon-C4 is semi-synthetic, one may question whether the generated queries are realistic enough for model evaluation. To complement it, we build Reddit-Movies based on the \texttt{reddit\_movie\_large\_v1} dataset~\cite{he23large}, which collects real Reddit discussions about movies and links each movie to its IMDb ID. Specifically, we keep only posts where users explicitly ask for movie suggestions (with the label \texttt{is\_seeker=True}) that have at least one valid response with more than 20 upvotes, which indicates a believable answer. 
For item metadata, we use the movie title and release year. The final dataset contains more than 20,000 query-item pairs, split into training, validation, and test sets following the original data split~\cite{he23large}.
Note that only the test split is used for evaluation in our benchmark.

\section{Experiments}

\subsection{Experimental Setup}

We evaluate 11 top-performing LLMs from the widely used general text embedding benchmark MTEB~\cite{muennighoff2023mteb,enevoldsen2025mmteb}:
(1) open-source LLMs with $< 1B$ parameters (RoBERTa~\cite{liu2017roberta}, SimCSE~\cite{gao2021simcse}, Sentence-T5~\cite{ni2022sentencet5});
(2) open-source LLMs with $\ge 1B$ parameters (E5~\cite{wang2024e5}, SFT-Embedding~\cite{sfr_embedding}, GritLM~\cite{muennighoff2025gritlm}); and
(3) proprietary text embedding models from Google (\texttt{gemini-embedding-001}) and OpenAI (\texttt{text-embedding-3-large}).
Among these, we further benchmark Qwen3-Embedding models of varying sizes (0.6B, 4B, and 8B)~\cite{zhang2025qwen3emb} to better understand the scaling behavior of LLM-based semantic encoders. Please refer to \Cref{app:exp-details} for more implementation details.

\begin{figure}[!t]
  \centering
  \includegraphics[width=0.9\columnwidth]{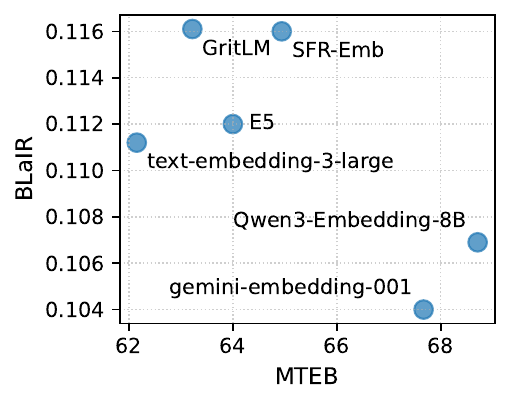}
  \caption{Performance correlation between MTEB (eng, v2) and BLaIR (avg. per task).}
  \label{fig:mteb-blair-corr}
\end{figure}

\subsection{Main Results}\label{sec:main_results} We present the benchmarking results in~\Cref{tab:main-results}, with detailed per-dataset results in~\Cref{app:full-results}. Several key observations emerge:

\textbf{(1) The ranking on \textsc{BLaIR} shows little correlation with MTEB.} By analyzing the Spearman correlation between the per-task average scores on \textsc{BLaIR} and MTEB (eng, v2)~\cite{enevoldsen2025mmteb}, we observe a correlation coefficient of $-0.476$ with $p=0.233$ (see~\Cref{fig:mteb-blair-corr}). This indicates that the performance of LLMs on general text embedding benchmarks does not necessarily reflect their effectiveness as semantic encoders for recommendation tasks. We attribute this to three possible factors. First, MTEB includes tasks such as clustering and summarization that are not directly aligned with the objectives of recommendation systems. Second, representations in recommendation tasks are expected to be highly discriminative~\cite{hou2022unisrec,sheng2024language}, a property not emphasized in MTEB. Third, as competition on MTEB intensifies, models may become overfitted to its specific tasks, resulting in reduced generalization to other domains.

\textbf{(2) The scaling behavior of semantic encoders weakens when downstream tasks become complex.}
According to scaling laws, larger models typically achieve better performance~\cite{kaplan2020scaling}, yet this trend is not consistently observed in our benchmark. We ensure a fair comparison by controlling the downstream model parameters across different LLM-based semantic encoders, using adaptors before feeding the representations into the downstream models. In tasks with simple downstream architectures, such as collaborative filtering (a single linear layer; see~\Cref{sec:task_def}), larger LLMs generally yield higher performance. However, in more complex settings like sequential recommendation, where the downstream model involves a Transformer decoder, the scaling trend becomes less pronounced. For instance, models in the Qwen3-Embedding family show comparable performance across different sizes on the sequential recommendation task. A discussion on this counterintuitive observation is provided in~\Cref{sec:discussion}.

\textbf{(3) The semi-synthetic dataset Amazon-C4 exhibits a strong linear correlation with the real-world dataset Reddit-Movie on the complex-query product search task.} By computing the Pearson correlation coefficient between the NDCG@100 scores of all evaluated models on Amazon-C4 and Reddit-Movie (\Cref{tab:prod-search-details}), we find a strong linearly positive correlation of $0.94$ ($p<0.01$). This indicates that Amazon-C4, despite being semi-synthetic and covering domains different from that in real-world complex queries, effectively captures essential characteristics of such queries. These results demonstrate that Amazon-C4 serves as a reliable proxy for evaluating (and potentially training) LLMs on complex-query product search tasks.

\textbf{(4) \texttt{text-emb-3-large} demonstrates unexpectedly strong generalization on \textsc{BLaIR}.} Developed by a leading company, \texttt{text-emb-3-large} is one of the most widely used text embedding models. Despite ranking only 42nd (as of Oct 6, AoE) on MTEB (English, v2), which is relatively low compared to other models evaluated in this study, raising doubts about its overall capability, it exhibits remarkably good performance across diverse datasets and tasks in \textsc{BLaIR}. Although its average performance is not the highest, its Borda Count ranking reveals consistently strong results across different scenarios. These observations suggest that \texttt{text-emb-3-large} possesses notable generalization ability and further indicate that evaluating LLM generalization remains an open and non-trivial challenge.

\begin{figure}[t]
    \centering
    \includegraphics[width=0.9\columnwidth]{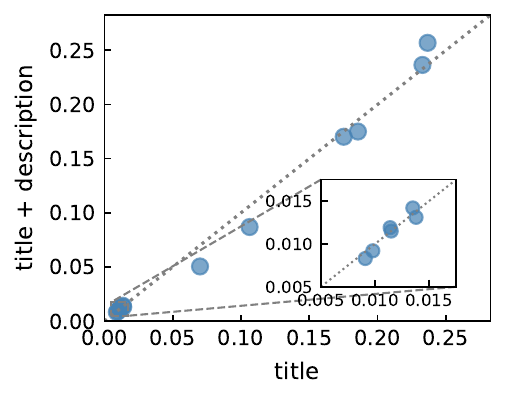}
    \caption{Comparison of performance when using different item metadata. 
    Each point represents a model on a specific dataset. }
    \label{fig:more_feat}
\end{figure}

\begin{table}[t]
    \small
    \centering
    \caption{Comparison between adaptors using matryoshka representation learning (MRL) and principal component analysis (PCA).}
    \label{tab:pca_mrl}
    \resizebox{\columnwidth}{!}{
\setlength{\tabcolsep}{1mm}{
    \begin{tabular}{lccc}
        \toprule
        \textbf{Model} & \textbf{Adaptor} & \textbf{Seq. Rec.} & \textbf{Col. Fil.} \\
        \midrule
        \multirow{2}{*}{\texttt{Qwen3-Embedding-8B}} & PCA & \textbf{0.0415} & 0.0362 \\
         & MRL & 0.0359 & \textbf{0.0392} \\
        \midrule
        \multirow{2}{*}{\texttt{gemini-embedding-001}} & PCA & \textbf{0.0434} & \textbf{0.0355} \\
         & MRL & 0.0384 & 0.0313 \\
        \midrule
        \multirow{2}{*}{\texttt{text-embedding-3-large}} & PCA & \textbf{0.0440} & 0.0366 \\
         & MRL & 0.0383 & \textbf{0.0379} \\
        \bottomrule
    \end{tabular}
    }
    }
\end{table}

\subsection{Performance \wrt Item Metadata}\label{sec:more_feat}
Although numerous item metadata attributes are available for use as item features in the collected Amazon Reviews 2023 dataset, in \textsc{BLaIR} we include only the item title, as it is the most universally available attribute, providing high information density in a concise and easily identifiable form. In this section, we investigate whether incorporating additional metadata can further enhance the performance of LLM-based semantic encoders. Specifically, we conduct experiments using both the item title and description as input features, and compare the results with those obtained using the title alone. We evaluate three representative models (\texttt{text-emb-3-large}, \texttt{Qwen3-Emb-8B}, and \texttt{SimCSE}) across four datasets: Games (sequential recommendation), Games (collaborative filtering), Amazon-C4, and ESCI.

The results in~\Cref{fig:more_feat} show that incorporating both the title and description does not consistently improve performance relative to using only the title. We attribute this to two factors. First, longer text inputs may introduce additional noise, making it harder for the model to extract relevant information.
Second, LLM world knowledge may already capture some of the information provided in item descriptions. These observations suggest that effectively leveraging rich, heterogeneous, and noisy item metadata in semantic encoders remains a challenging open problem.

\subsection{Performance \wrt Adaptor Design}

In our experiments, to ensure fair comparisons across different LLM-based semantic encoders, we employ adaptors to align their representation dimensions. We use principal component analysis (PCA) with whitening as the default adaptor due to its simplicity and efficiency~\cite{huang2021whiteningbert,su2021whitening}. Meanwhile, several LLMs are capable of directly producing low-dimensional representations through matryoshka representation learning (MRL)~\cite{kusupati2022matryoshka}. In this section, we compare the performance of PCA and MRL on three representative models (\texttt{Qwen3-Embedding-8B}, \texttt{gemini-embedding-001}, and \texttt{text-embedding-3-large}) across all recommendation datasets that require an adaptor.

As shown in~\Cref{tab:pca_mrl}, for tasks involving complex downstream models such as sequential recommendation, PCA outperforms MRL, likely because whitening encourages more distinguishable representations. In contrast, for tasks with simpler architectures like collaborative filtering, MRL tends to achieve better results, possibly due to its ability to preserve task-relevant information in low-dimensional spaces. These findings suggest that the choice of adaptor should be guided by the complexity of the downstream task and model architecture.

\section{Conclusion}

In this work, we introduced \textsc{BLaIR}, a comprehensive benchmark for evaluating LLMs as semantic encoders in recommendation scenarios. We contributed (1) a new large-scale Amazon Reviews 2023 dataset, (2) a unified benchmark spanning sequential recommendation, collaborative filtering, and product search, and (3) a novel complex-query product search task with both semi-synthetic and real-world evaluation datasets. Our experiments with 11 leading LLMs revealed several key insights: (1) LLM rankings on \textsc{BLaIR} show little correlation with general text embedding benchmarks like MTEB, indicating that the role of semantic encoders poses unique challenges. (2) Scaling laws apply to semantic encoders even when downstream models are parameter-controlled, though the effect weakens with task complexity increases. (3) Our semi-synthetic Amazon-C4 dataset shows strong correlation with real-world complex queries, indicating that it can serve as a reliable evaluation proxy and may also benefit the training of LLMs for complex-query tasks.

\clearpage

\section*{Acknowledgement}

This work is partially supported by NSF IIS-2432486.

\section*{Limitations}

While \textsc{BLaIR} provides a comprehensive evaluation framework for LLMs as semantic encoders in recommendation, several limitations should be acknowledged. First, our current benchmark focuses exclusively on English-language data. Given the global nature of e-commerce and recommender systems, multilingual evaluation would be valuable for assessing the cross-lingual capabilities of semantic encoders. Second, due to computational budget constraints, our experiments are limited to 11 LLMs and a subset of categories from the Amazon Reviews 2023 dataset. A more extensive evaluation covering additional state-of-the-art models and a broader range of categories would provide deeper insights into the generalizability of our findings.

\section*{Ethical Considerations}

Our work prioritizes user privacy and data ethics in several ways. For the Amazon Reviews 2023 dataset, we exclusively collect publicly available information that users have explicitly chosen to share. We do not include any content that users have opted to keep private or restricted. Importantly, our dataset does not contain user metadata to prevent potential misuse for user profiling or privacy violations. The dataset focuses solely on publicly visible product reviews and item metadata, which are essential for recommendation research while minimizing privacy risks.

Regarding the \textsc{BLaIR} benchmark, all evaluation tasks are constructed using publicly released datasets that have been widely adopted by the research community. We ensure that our benchmark adheres to the original data collection and usage policies of these datasets. While recommender systems can potentially reinforce biases or create filter bubbles, our benchmark is designed to evaluate semantic encoding capabilities rather than deployment-ready recommendation systems. We encourage researchers using \textsc{BLaIR} to consider fairness, transparency, and potential societal impacts when developing real-world applications based on insights from this benchmark.

\bibliography{custom}

\clearpage

\appendix

\begin{center}
    {\Large Appendices}
\end{center}

\providecommand{\uline}[1]{\underline{#1}}

\begingroup
\setlength{\tabcolsep}{1.6mm}
\renewcommand{\arraystretch}{0.95}

\begin{table*}[!htbp]
\small
\centering
\caption{Sequential Recommendation (NDCG@10) detailed results. Best per column in \textbf{bold}, second best \uline{underlined}.}
\label{tab:seq-rec-details}
\scriptsize
\begin{tabular}{lrrrrr}
\toprule
\textbf{Model / Datasets} & \textbf{All\_Beauty} & \textbf{Video\_Games} & \textbf{Baby\_Products} & \textbf{ML1M} & \textbf{Yelp} \\
\midrule
\rowcolor{gray!20}\multicolumn{6}{c}{\textbf{\textit{Open-Source Models (< 1B parameters)}}} \\
\midrule
\texttt{FacebookAI/roberta-large} & 0.0177 & 0.0113 & 0.0070 & 0.1215 & 0.0391 \\
\texttt{Qwen/Qwen3-Embedding-0.6B} & 0.0224 & 0.0126 & 0.0072 & 0.1266 & 0.0385 \\
\texttt{princeton-nlp/sup-simcse-roberta-large} & 0.0232 & 0.0114 & 0.0069 & 0.1312 & 0.0404 \\
\texttt{sentence-transformers/sentence-t5-large} & 0.0212 & 0.0125 & 0.0075 & 0.1275 & 0.0405 \\
\midrule
\rowcolor{gray!20}\multicolumn{6}{c}{\textbf{\textit{Open-Source Models ($\ge$ 1B parameters)}}} \\
\midrule
\texttt{Qwen/Qwen3-Embedding-4B} & 0.0212 & 0.0127 & 0.0075 & 0.1276 & 0.0392 \\
\texttt{Qwen/Qwen3-Embedding-8B} & 0.0216 & \textbf{0.0138} & 0.0072 & 0.1248 & 0.0400 \\
\texttt{Salesforce/SFR-Embedding-Mistral} & 0.0231 & 0.0131 & \uline{0.0076} & 0.1311 & 0.0416 \\
\texttt{intfloat/e5-mistral-7b-instruct} & \uline{0.0238} & \uline{0.0136} & 0.0072 & \uline{0.1324} & 0.0398 \\
\texttt{GritLM/GritLM-7B} & 0.0216 & 0.0133 & 0.0074 & 0.1321 & \textbf{0.0426} \\
\midrule
\rowcolor{gray!20}\multicolumn{6}{c}{\textbf{\textit{Proprietary Models}}} \\
\midrule
\texttt{gemini-embedding-001} (Google) & \textbf{0.0241} & \uline{0.0136} & 0.0073 & 0.1297 & \uline{0.0421} \\
\texttt{text-embedding-3-large} (OpenAI) & 0.0237 & 0.0135 & \textbf{0.0078} & \textbf{0.1330} & 0.0418 \\
\bottomrule
\end{tabular}
\end{table*}

\begin{table*}[!htbp]
\small
\centering
\caption{Collaborative Filtering (NDCG@20) detailed results. Best per column in \textbf{bold}, second best \uline{underlined}.}
\label{tab:cf-details}
\scriptsize
\begin{tabular}{lrrrrrr}
\toprule
\textbf{Model / Datasets} & \textbf{All\_Beauty} & \textbf{Video\_Games} & \textbf{Baby\_Products} & \textbf{Book\_Crossing} & \textbf{ML1M} & \textbf{Yelp} \\
\midrule
\rowcolor{gray!20}\multicolumn{7}{c}{\textbf{\textit{Open-Source Models (< 1B parameters)}}} \\
\midrule
\texttt{FacebookAI/roberta-large} & 0.0183 & 0.0087 & 0.0037 & 0.0338 & 0.0658 & \textbf{0.0310} \\
\texttt{Qwen/Qwen3-Embedding-0.6B} & 0.0220 & 0.0090 & 0.0047 & 0.0296 & 0.0713 & 0.0280 \\
\texttt{princeton-nlp/sup-simcse-roberta-large} & 0.0205 & 0.0091 & 0.0040 & 0.0285 & 0.0676 & 0.0295 \\
\texttt{sentence-transformers/sentence-t5-large} & 0.0202 & 0.0101 & 0.0043 & 0.0329 & 0.0869 & 0.0280 \\
\midrule
\rowcolor{gray!20}\multicolumn{7}{c}{\textbf{\textit{Open-Source Models ($\ge$ 1B parameters)}}} \\
\midrule
\texttt{Qwen/Qwen3-Embedding-4B} & 0.0219 & 0.0099 & 0.0041 & 0.0406 & 0.1047 & 0.0285 \\
\texttt{Qwen/Qwen3-Embedding-8B} & 0.0226 & 0.0098 & 0.0042 & \uline{0.0427} & 0.1097 & 0.0284 \\
\texttt{Salesforce/SFR-Embedding-Mistral} & 0.0236 & 0.0107 & 0.0049 & 0.0416 & 0.1142 & 0.0284 \\
\texttt{intfloat/e5-mistral-7b-instruct} & 0.0232 & \uline{0.0110} & 0.0049 & 0.0421 & \uline{0.1162} & 0.0287 \\
\texttt{GritLM/GritLM-7B} & 0.0226 & \uline{0.0110} & 0.0046 & 0.0425 & \textbf{0.1216} & 0.0287 \\
\midrule
\rowcolor{gray!20}\multicolumn{7}{c}{\textbf{\textit{Proprietary Models}}} \\
\midrule
\texttt{gemini-embedding-001} (Google) & \textbf{0.0242} & 0.0108 & \textbf{0.0051} & 0.0371 & 0.1054 & \uline{0.0304} \\
\texttt{text-embedding-3-large} (OpenAI) & \uline{0.0238} & \textbf{0.0115} & \uline{0.0050} & \textbf{0.0436} & 0.1071 & 0.0286 \\
\bottomrule
\end{tabular}
\end{table*}

\begin{table*}[!htbp]
\small
\centering
\caption{Product Search (NDCG@100) detailed results. Best per column in \textbf{bold}, second best \uline{underlined}.}
\label{tab:prod-search-details}
\scriptsize
\begin{tabular}{lrrr}
\toprule
\textbf{Model / Datasets} & \textbf{Amazon-C4} & \textbf{ESCI} & \textbf{Reddit-Movies} \\
\midrule
\rowcolor{gray!20}\multicolumn{4}{c}{\textbf{\textit{Open-Source Models (< 1B parameters)}}} \\
\midrule
\texttt{FacebookAI/roberta-large} & 0.0001 & 0.0096 & 0.0000 \\
\texttt{Qwen/Qwen3-Embedding-0.6B} & 0.1288 & 0.1876 & 0.0211 \\
\texttt{princeton-nlp/sup-simcse-roberta-large} & 0.0700 & 0.1063 & 0.0078 \\
\texttt{sentence-transformers/sentence-t5-large} & 0.1219 & 0.1691 & 0.0360 \\
\midrule
\rowcolor{gray!20}\multicolumn{4}{c}{\textbf{\textit{Open-Source Models ($\ge$ 1B parameters)}}} \\
\midrule
\texttt{Qwen/Qwen3-Embedding-4B} & 0.1663 & 0.2258 & 0.0576 \\
\texttt{Qwen/Qwen3-Embedding-8B} & 0.1752 & 0.2328 & 0.0591 \\
\texttt{Salesforce/SFR-Embedding-Mistral} & 0.1836 & \textbf{0.2560} & \uline{0.0710} \\
\texttt{intfloat/e5-mistral-7b-instruct} & 0.1759 & 0.2437 & 0.0705 \\
\texttt{GritLM/GritLM-7B} & \uline{0.1845} & \uline{0.2537} & \textbf{0.0734} \\
\midrule
\rowcolor{gray!20}\multicolumn{4}{c}{\textbf{\textit{Proprietary Models}}} \\
\midrule
\texttt{gemini-embedding-001} (Google) & 0.1701 & 0.2233 & 0.0578 \\
\texttt{text-embedding-3-large} (OpenAI) & \textbf{0.1856} & 0.2366 & 0.0699 \\
\bottomrule
\end{tabular}
\end{table*}

\endgroup

\section{Benchmarking Toolkit}

To facilitate reproducibility and simplify the integration of new models and datasets, we have developed a benchmarking toolkit available at \url{https://github.com/hyp1231/BLaIR-Bench}. The toolkit includes implementations of all datasets, tasks, and evaluation metrics used in this study, along with scripts for model training and evaluation. We welcome community contributions to further extend the benchmark.

\section{Dataset Collection Details}

The Amazon Reviews 2023 dataset is collected using a large-scale, user-centric pipeline. We first sample user identifiers from public Amazon pages and then iteratively collect all review pages associated with each user. This user-level design ensures completeness of individual review histories, which is particularly beneficial for recommendation and user modeling research.

However, the review coverage for a given item may be incomplete if some users are not included in our user pool. Additionally, due to the temporal span of the collecting process, recently posted reviews might be missing even before the dataset cut-off date.

\begin{table*}[t]
\centering
\small
\begin{tabular}{p{0.18\textwidth} p{0.76\textwidth}}
\toprule
\textbf{User query} & ``Over the Top Bonkers Action Movies?. I recently saw the new \textit{Suicide Squad} movie and loved it. It had all the qualities I like in an action movie: violence, shock factor, comedy, and overall bizarreness. What are some other action movies with these qualities? \textit{Deadpool}, \textit{Kick-Ass}, \textit{Turbo Kid}, \textit{Tropic Thunder}, and \textit{Drive Angry} are some of my other favorites.'' \\
\midrule
\textbf{Relevant items} & \textit{Nobody} (2021), \textit{Hardcore Henry} (2015), \textit{Slow Moe} (2010), \textit{Dredd} (2012) \\
\midrule
\textbf{Top-3 retrieved by GritLM-7B} & \textit{Ninjas vs. Zombies} (2008), \textit{Cowboys vs. Zombies} (2014), \textit{Cowboys vs. Vampires} (2010) \\
\midrule
\textbf{Analysis} & The retrieved movies appear to match the query only at a superficial level. For example, they contain unusual or fantastical entities such as zombies, ninjas, and vampires, which may loosely correlate with ``bizarreness.'' However, they fail to capture the core intent of the query, namely over-the-top action movies that combine stylized violence, dark humor, and mainstream appeal. This example suggests that current semantic encoders still struggle to model complex intent beyond shallow lexical or thematic similarity. \\
\bottomrule
\end{tabular}
\caption{A failure case of the top-performing semantic encoder GritLM-7B on complex-query product search in the Reddit-Movies dataset.}
\label{tab:case_reddit_movie}
\end{table*}

\section{Experimental Setup}\label{app:exp-details}

\subsection{Data Split and Preprocessing}\label{app:data-split}

\paragraph{Sequential Recommendation.}
We group all user-item interactions by user ID and sort them by timestamp. For each user, we truncate their interaction history to the most recent 50 items (200 for ML-1M). Dataset-specific preprocessing steps are as follows:
\begin{itemize}
    \item \textbf{Amazon Reviews 2023.} To preserve the natural distribution, we do not filter out any users or items based on interaction counts. Since sequential recommendation is time-sensitive, we split the dataset by time. To be specific, we determined two cut-off timestamps \texttt{1628643414042} and \texttt{1658002729837} that split all the reviews in Amazon Reviews 2023 in a ratio of $8:1:1$. These timestamps are shared across the three categories: All Beauty, Video Games, and Baby Products.
    \item \textbf{Movielens-1M.} We iteratively remove users and items with fewer than 20 interactions until convergence, then apply the leave-last-out strategy: the last two interactions per user are used for validation and testing, while all previous interactions are used for training.
    \item \textbf{Yelp.} We iteratively remove users and items with fewer than 5 interactions until convergence and apply the same leave-last-out strategy as ML-1M. Using the 2020 competition dataset, we keep only interactions from 2019.
\end{itemize}

\paragraph{Collaborative Filtering.} Following \citet{sheng2024language}, we discard timestamps and split user-item interactions into training, validation, and test sets with a 4:3:3 ratio. For ML-1M and Book-Crossing, users and items with fewer than 20 interactions are iteratively removed. For Yelp, the threshold is 5. No filtering is applied to the three Amazon Reviews 2023 categories.

\paragraph{Short-Query Product Search (ESCI).}
We use the ESCI dataset~\cite{reddy2022esci} and retain only \textit{<query, item>} pairs with the label ``Exact''. We adopt the small version due to its higher-quality queries, filter out non-English queries, and keep only those whose timestamps fall within our predefined test period (after \texttt{1658002729837}). Each item ID (\texttt{ASIN}) is linked to its metadata in Amazon Reviews 2023. We construct a large multi-category candidate pool by sampling 50 in-domain items (from the same category) for each ground-truth pair. Ranking is then performed over all multi-domain candidates for each query.

\paragraph{Complex-Query Product Search.}
\begin{itemize}
    \item \textbf{Amazon-C4.} The data synthesizing process is described in~\Cref{sec:new_task}. After that, we follow the same multi-category candidate sampling strategy as ESCI, sampling 50 in-domain items for each ground-truth pair. Ranking is computed over both in-domain and cross-domain candidates.
    \item \textbf{Reddit-Movie.} We use the \texttt{reddit\_movie\_large\_v1} dataset~\cite{he23large}, which contains multi-turn conversations where users seek movie recommendations. We retain only conversations with the label \texttt{is\_seeker=True} (indicating the user is seeking movie recommendations) and apply several filtering steps:
(1) contexts must include at least one valid seeker turn;
(2) recommendations must have at least 20 upvotes;
(3) duplicate entries are removed using MD5 hashes of concatenated text and context;
(4) samples must contain valid movie IDs matched with metadata.
The dataset is split following the original partition, and each \textit{<query, item>} pair represents a complex query and a recommended movie.
\end{itemize}

\subsection{Item Metadata Processing}
\begin{itemize}
    \item \textbf{Amazon Reviews 2023.} For all Amazon-related datasets (Beauty, Games, Baby, Amazon-C4, and ESCI), we use product titles as item metadata. Although we conducted a small-scale experiment using concatenated titles and descriptions (\Cref{sec:more_feat}), the inclusion of descriptions did not consistently improve performance. Therefore, we use only titles for simplicity and consistency.
    \item \textbf{Movielens-1M.} Movie titles and release years are used as metadata.
    \item \textbf{Yelp.} We concatenate each business's `name', `categories', `stars', `review\_count', `city', `state', `attributes', and `hours' fields.
    \item \textbf{Book-Crossing.} We use book titles as metadata.
    \item \textbf{Reddit-Movie.} We use movie titles and release years as metadata.
\end{itemize}

\subsection{Evaluation Metrics}
For each dataset, we use the normalized discounted cumulative gain (NDCG) as the primary metric, with task-specific cutoffs: NDCG@10 for sequential recommendation, NDCG@20 for collaborative filtering, and NDCG@100 for product search.

To summarize results across datasets, we report three aggregated metrics in~\Cref{tab:main-results}. First, to alleviate the bias in the number of datasets per task and the differences in metric scales, we follow MMTEB~\cite{enevoldsen2025mmteb} and adopt the Borda Count~\cite{colombo2022borda} as the primary ranking criterion. Second, for intuitive comparison, we also report the average across all datasets (Avg. Overall) and the average after first averaging within each (sub)task (Avg. per Task).

\subsection{Evaluation Pipeline}

For tasks requiring model training (sequential recommendation and collaborative filtering), we search over three learning rates ($\{1\times 10^{-3}, 3\times 10^{-4}, 1\times 10^{-4}\}$) and select the checkpoint with the best validation performance for final testing.

For product search, evaluation is conducted in a zero-shot manner without training, following each LLM's recommended encoding procedure; for models that suggest a prompt when encoding queries, we use ``\textit{Given a user query describing a need, retrieve relevant entities (such as products, services, or media items) that satisfy the intent and constraints.}''.

\section{Detailed Benchmarking Results}\label{app:full-results}

The results for each task are summarized in~\Cref{tab:main-results}. Below, we provide detailed results for all evaluated LLMs across individual datasets. The results for sequential recommendation, collaborative filtering, and product search are presented in~\Cref{tab:seq-rec-details}, \Cref{tab:cf-details}, and \Cref{tab:prod-search-details}, respectively.

\section{Discussion}\label{sec:discussion}

\paragraph{Why is the correlation between BLaIR and MTEB results weak?} We hypothesize that building a strong semantic encoder requires three key capabilities that may not be simultaneously emphasized in existing embedding benchmarks:
\begin{itemize}[leftmargin=1em]
    \item The ability to accurately model semantic similarity, which is essential for tasks such as collaborative filtering and short-query product search.
    \item The ability to produce distinguishable and informative feature representations, which is important for tasks like sequential recommendation.
    \item Strong instruction-following and reasoning abilities, which are crucial for handling complex-query product search.
\end{itemize}

\paragraph{Why scaling semantic encoders yields limited gains on complex tasks?} As observed in~\Cref{sec:main_results}, in tasks with complex downstream architectures, such as sequential recommendation (a Transformer decoder), the scaling trend becomes less
pronounced than tasks with simple downstream model architectures like collaborative filtering (a simple linear layer).  Our current hypotheses are:
\begin{itemize}[leftmargin=1em]
    \item \emph{Differences in the nature of the tasks}: The representational capabilities required differ between tasks. Collaborative filtering is primarily about encoding similarity. A stronger semantic encoder can represent item similarities in a richer and more accurate manner, which directly translates into performance gains in collaborative filtering tasks as the encoder scales up. For example, \citet{sheng2024language} found that stronger semantic encoders yield representations whose distribution is more closely aligned with those learned from pure interaction data. Sequential recommendation is essentially a sequence modeling problem that requires memorizing and reproducing behavior patterns. To achieve this, item representations may require strong discriminative capability (akin to functioning as unique indices). Representations produced by even a smaller semantic encoder might already be sufficiently distinguishable for a complex downstream Transformer model. Consequently, scaling the semantic encoder does not significantly enhance discriminability from the perspective of the downstream model, leading to smaller gains compared to collaborative filtering. A similar observation was reported by~\citet{hou2022unisrec}, who showed that treatments such as whitening enhance the discriminative power of item representations and improve sequential recommendation performance.
    \item \emph{Sequentially dependent neural network systems and scaling behaviors}: Our setup functions as a two-stage system: a first-stage semantic encoder whose output is consumed by a second-stage model (a linear layer for collaborative filtering, or a Transformer decoder for sequential recommendation). We hypothesize that in such multi-stage systems, scaling the later-stage modules may diminish the impact of scaling earlier-stage modules. While there is limited systematic work on scaling laws in sequentially dependent systems, our findings may suggest that where we allocate additional parameters may matter as much as how many parameters we add.
\end{itemize}

\section{Case Study: Failure on Complex-Query Product Search}
\label{app:case_study_reddit_movie}

Table~\ref{tab:case_reddit_movie} presents a failure case from the Reddit-Movies dataset, which requires understanding a complex user query with multiple implicit preferences. 
This example highlights headroom for improvement in complex-query product search. In particular, even the best-performing semantic encoder on this dataset, GritLM-7B, only achieves an NDCG@100 of 0.0734 over a candidate set of 50k movies, suggesting that surface-level semantic matching is far from sufficient. Future improvements may require stronger intent modeling, explicit reasoning, or the incorporation of additional ranking signals such as popularity.

\end{document}